\documentclass[sigconf, screen, nonacm]{acmart}

\AtBeginDocument{%
  }

\usepackage{hyperref}
\usepackage{cleveref}
\usepackage{graphicx}
\usepackage{svg}
\usepackage{subcaption}
\begin{document}

\title{Rethinking Compute Substrates for 3D-Stacked Near-Memory LLM Decoding: Microarchitecture–Scheduling Co-Design}
\author{Chenyang Ai}
\authornote{For any questions regarding this manuscript, please contact me.}
\affiliation{%
  \institution{University of Edinburgh}
  \city{Edinburgh}
  \country{United Kingdom}}
\email{C.Ai-4@sms.ed.ac.uk}

\author{Yixing Zhang}
\affiliation{%
  \institution{Peking University}
  \city{Beijing}
  \country{China}}
\email{2401213080@stu.pku.edu.cn}

\author{Haoran Wu}
\affiliation{%
  \institution{University of Cambridge}
  \city{Cambridge}
  \country{United Kingdom}}
\email{hw691@cam.ac.uk}

\author{Yudong Pan}
\affiliation{%
  \institution{University of Chinese Academy of Sciences}
  \city{Beijing}
  \country{China}}
\email{panyudong23@mails.ucas.ac.cn}

\author{Lechuan Zhao}
\affiliation{%
  \institution{Peking University}
  \city{Beijing}
  \country{China}}
\email{lczhao@stu.pku.edu.cn}

\author{Wenhui OU}
\affiliation{%
  \institution{The Hong Kong University of Science and Technology}
  \city{Hong Kong}
  \country{China}}
\email{wouab@connect.ust.hk}




\begin{abstract}

Large language model (LLM) decoding is a major inference bottleneck because its low arithmetic intensity makes performance highly sensitive to memory bandwidth. 3D-stacked near-memory processing (NMP) provides substantially higher local memory bandwidth than conventional off-chip interfaces, making it a promising substrate for decode acceleration. However, our analysis shows that this bandwidth advantage also shifts many decode operators on 3D-stacked NMP back into the compute-bound regime. Under the tight area budget of the logic die, the design of the compute substrate itself therefore becomes a first-order challenge.

Therefore, we rethink the compute microarchitecture of prior 3D-stacked NMP designs. First, we replace prior MAC tree-based compute units with a more area-efficient systolic array, and we further observe that decode operators exhibit substantial shape diversity, making reconfigurability in both systolic array shape and dataflow essential for sustaining high utilization. Building on this insight, we continue to exploit two key opportunities: the high local memory bandwidth reduces the need for large on-chip buffers, and the existing vector core, originally designed to handle auxiliary tensor computations, already provides much of the control logic and multi-ported buffering required for fine-grained flexibility for systolic array, allowing us to unify the two structures in a highly area-efficient manner. Based on these insights, we present the first compute microarchitecture tailored to 3D-stacked NMP LLM decoding, explicitly designed to satisfy the joint requirements of low area cost, high-bandwidth operation, and fine-grained reconfigurability.

To scale the design across multiple cores on one logic die, we further propose an operator-aware scheduling framework that combines spatial and spatio-temporal partitioning for LLM decode operators. Compared with Stratum, our design achieves an average 2.91× speedup and 2.40× higher energy efficiency across both dense and MoE models.

\end{abstract}


\settopmatter{printacmref=false}
\maketitle

\section{INTRODUCTION}

In recent years, Large Language Models (LLMs) have advanced rapidly. LLM inference is typically divided into two phases, namely prefill and decode. Among these, the decode phase is particularly critical, as it directly determines user-perceived responsiveness. Decode proceeds in a token-by-token manner, offering limited weight reuse and leading to low arithmetic intensity. As a result, for both dense and MoE models, end-to-end performance is primarily constrained by off-chip memory bandwidth. Consequently, optimizing LLM decode has become a central focus of recent research~\cite{agrawal2023sarathi,li2024llm,zhou2024survey}.




In recent years, 3D-stacked DRAM-based near-memory processing (NMP) has emerged as a promising architectural paradigm for accelerating LLM decode~\cite{he2025tasa,pan2025stratum,yun2024duplex,li2026helios,han2025near}. These designs integrate compute logic directly within the memory stack, typically on a logic-based die beneath multiple stacked DRAM layers. By leveraging dense vertical interconnects, such as Through-Silicon Vias (TSVs)~\cite{van2016small} or hybrid bonding~\cite{chen20203d}, NMP architectures enable substantially higher internal memory bandwidth between the logic die and DRAM layers compared to conventional off-chip interfaces. This high-bandwidth, low-latency data access is particularly beneficial for memory-bound operations during decode. Prior work further demonstrates that NMP architectures can operate in a heterogeneous manner alongside traditional xPUs, such as GPUs~\cite{choquette2023h100} or TPUs~\cite{jouppi2025ironwood_hotchips}, enabling cooperative execution that improves end-to-end LLM inference performance~\cite{yun2024duplex,pan2025stratum}.

\begin{figure}[t]
    \centering
\includegraphics[width=\linewidth,trim=0.7cm 0.7cm 0.3cm 0.3cm,clip]{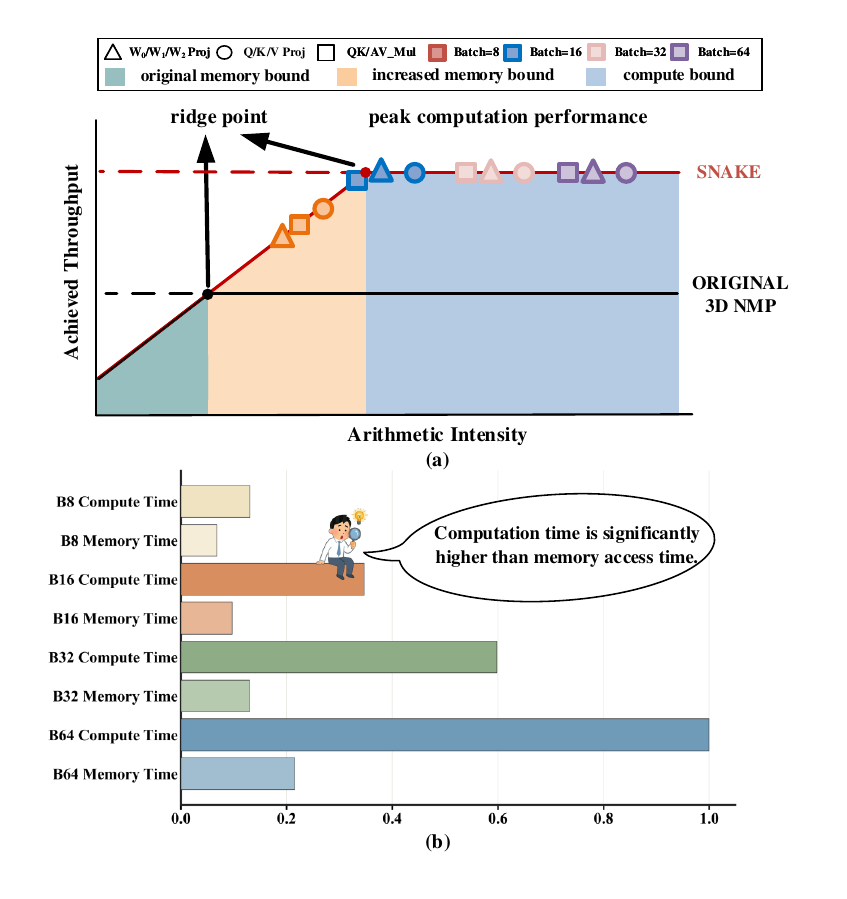}
    \caption{(a) Roofline Analysis of LLM Decode Operators on 3D-Stacked NMP (b) Memory-Side Execution Analysis of Decode Operators in Stratum}
    \Description{Limitations of existing 3D-stacked NMP architectures for LLM decoding. Despite high internal memory bandwidth, both Duplex and Stratum remain compute-limited, resulting in under-utilized memory bandwidth and limited latency hiding.}
    \label{fig:roofline}
\end{figure}

However, in the 3D NMP setting, many decode tensor operators shift from being memory-bound back to compute-bound. Since decode performance remains primarily dominated by tensor operators, we focus our analysis on matrix multiplication and its associated memory-access behavior from a roofline perspective. As shown in Figure~\ref{fig:roofline} (a), existing 3D NMP architectures have not scaled on-stack compute capability commensurately with the rapid growth in internal memory bandwidth. Taking Duplex~\cite{yun2024duplex} and Stratum~\cite{pan2025stratum} as representative 3D-stacked NMP architectures, their compute-to-memory-bandwidth ratio only reaches 8 and 3.7–6.7 FLOPs/Byte, respectively, while decode operators are often required to support batch sizes of up to 64 in practical serving~\cite{pan2025stratum,li2026helios}, resulting in arithmetic intensity that frequently and significantly exceeds these hardware ridge points. As corroborating evidence in Figure~\ref{fig:roofline} (b), our reproduction of Stratum~\cite{pan2025stratum}, conducted on LLaMA3 models across varying batch sizes, shows that, even with double buffering to overlap data movement with execution, array compute time remains substantially longer than memory-supply time, indicating that the currently provisioned compute throughput significantly lags behind the available memory supply capability.

Intrinsically, in heterogeneous LLM inference systems, the compute-bound behavior of 3D NMP arises from operator scheduling. Mainstream xPUs typically exhibit ridge points on the order of hundreds of FLOP/Byte or higher~\cite{jouppi2025ironwood_hotchips,choquette2023h100}, making them naturally well suited for high–arithmetic-intensity prefill operators. Although techniques such as continuous batching and attention variants (e.g., GQA/MQA) can increase arithmetic intensity, decode operators still remain far below prefill workloads. Therefore, in heterogeneous systems, decode operators are preferentially assigned to 3D-stacked NMP, where even many medium-to-low arithmetic-intensity decode operators become compute-bound thanks to the high local memory bandwidth.




One of the fundamental limitations is the severely constrained area available to the NMP engine on the logic die. In practice, only a limited fraction of the logic die can be allocated to active compute logic. Stratum is a representative example: to maintain compatibility with the HBM3 xPU–DRAM interface, the logic die must reserve substantial area for the HBM3 PHY, DRAM peripherals, and power-delivery overhead, which significantly compresses the area budget available for active logic. Then, HBM logic dies remain physically compact—e.g., around 121 mm² in HBM3-class designs—because their footprint is constrained by advanced-package assembly and reliability requirements, a tight thermal envelope, and poor cost efficiency under die-area scaling~\cite{kim2024present,moon2023advanced,lau2022recent}. As a result, Stratum reports an active logic area of only about 76.63 mm². In contrast, the neighboring xPU typically exposes only a fraction of the memory bandwidth available on the NMP side, yet can devote an effective compute area that is an order of magnitude larger.  

Therefore, 3D-stacked NMP requires a different compute substrate design point: one that maximizes compute density under a severely constrained logic-die area budget. From the perspective of compute unit, MAC-Tree-based designs have been widely adopted in prior work~\cite{pan2025stratum,yun2024duplex,huang2025hd,li2025h2} as a mainstream solution. However, as their scale increases, the high-fanout operand buffer delivery paths and multi-stage reduction networks tend to incur substantial interconnect and control overhead, ultimately limiting area efficiency and scalability (detailed in Section ~\ref{bg}). In contrast, Systolic
Array (SA) organizes a large number of simple processing elements, each performing only basic multiply–accumulate operations, into a two-dimensional structure connected through regular nearest-neighbor links. By exploiting structured data movement, it enables high compute density and array-level data reuse within the fabric, and therefore typically offers superior area efficiency and energy efficiency. This makes it a more suitable compute substrate for area-constrained 3D NMP architectures. This architectural advantage is also borne out by our RTL implementations: under the same frequency and PE-level compute functionality, the MAC-Tree design requires 8.23× more area than SA.

However, directly adopting a conventional SA for LLM decode on 3D-stacked NMP is not straightforward. A closer examination of this setting reveals two key requirements. Operators mapped to a single core during decode exhibit substantial shape diversity, making it difficult for a fixed-shape array to sustain high utilization, especially when many of these operators have already become compute-bound. Moreover, the dimensional relationships among operators can also change the preferred systolic dataflow. A decode-oriented SA therefore must support reconfigurability in both array shape and dataflow.


Beyond the intrinsic area efficiency of SA, the 3D-stacked NMP setting also creates two architectural opportunities to further improve compute-area efficiency and thereby deploy more compute to alleviate bottlenecks. On the one hand, higher local memory bandwidth weakens the conventional need for large-capacity buffers, making it possible to increase the number of compute units by reducing buffer area. On the other hand, the auxiliary vector core already provides fine-grained control and flexible buffering, offering a useful foundation for exploring a unified systolic-vector compute substrate, and thus may enable the required reconfigurability at lower area overhead.

Based on these insights, we propose SNAKE, a reconfigurable SA tailored for LLM decode on 3D-stacked NMP, together with a co-designed multi-core scheduling framework. More broadly, architecture research has rarely studied multi-core scheduling for reconfigurable SA in a systematic way. We address this gap by exploring this scheduling space and applying it to high-bandwidth 3D-stacked NMP. The main contributions are as follows:
\begin{itemize}

\item We identify the compute-area bottleneck of LLM decode on current 3D-stacked NMP and derive from it the need for reconfigurable systolic execution in both array shape and dataflow, together with two architectural opportunities that enable the required flexibility while freeing area for additional compute units.

\item We propose SNAKE, a reconfigurable SA whose microarchitecture is optimized for 3D-stacked NMP, meeting the key requirements of LLM decode: high bandwidth utilization, low area overhead, and fine-grained reconfigurability. To achieve this, we introduce a systolic–vector architecture combined with SNAKE-like mapping for area efficiency and high utilization.

\item We introduce a multi-core scheduling framework, enabled by a lightweight on-chip interconnect, that efficiently exploits the abundant array resources and high local bandwidth of 3D-stacked NMP by aligning systolic dataflows with spatial and spatio-temporal partitioning.

\item Compared with state-of-the-art 3D NMP baseline, SNAKE achieves 4.00× higher compute-area efficiency, and delivers an average 2.90× speedup and 2.40× higher energy efficiency across diverse LLMs.

\end{itemize}

\section{BACKGROUND AND DESIGN CONTEXT}
\label{bg}
\textbf{MAC Tree and Systolic Array}
\begin{figure}[t]
    \centering
\includegraphics[width=\linewidth,trim=0 0cm 0 0cm,clip]{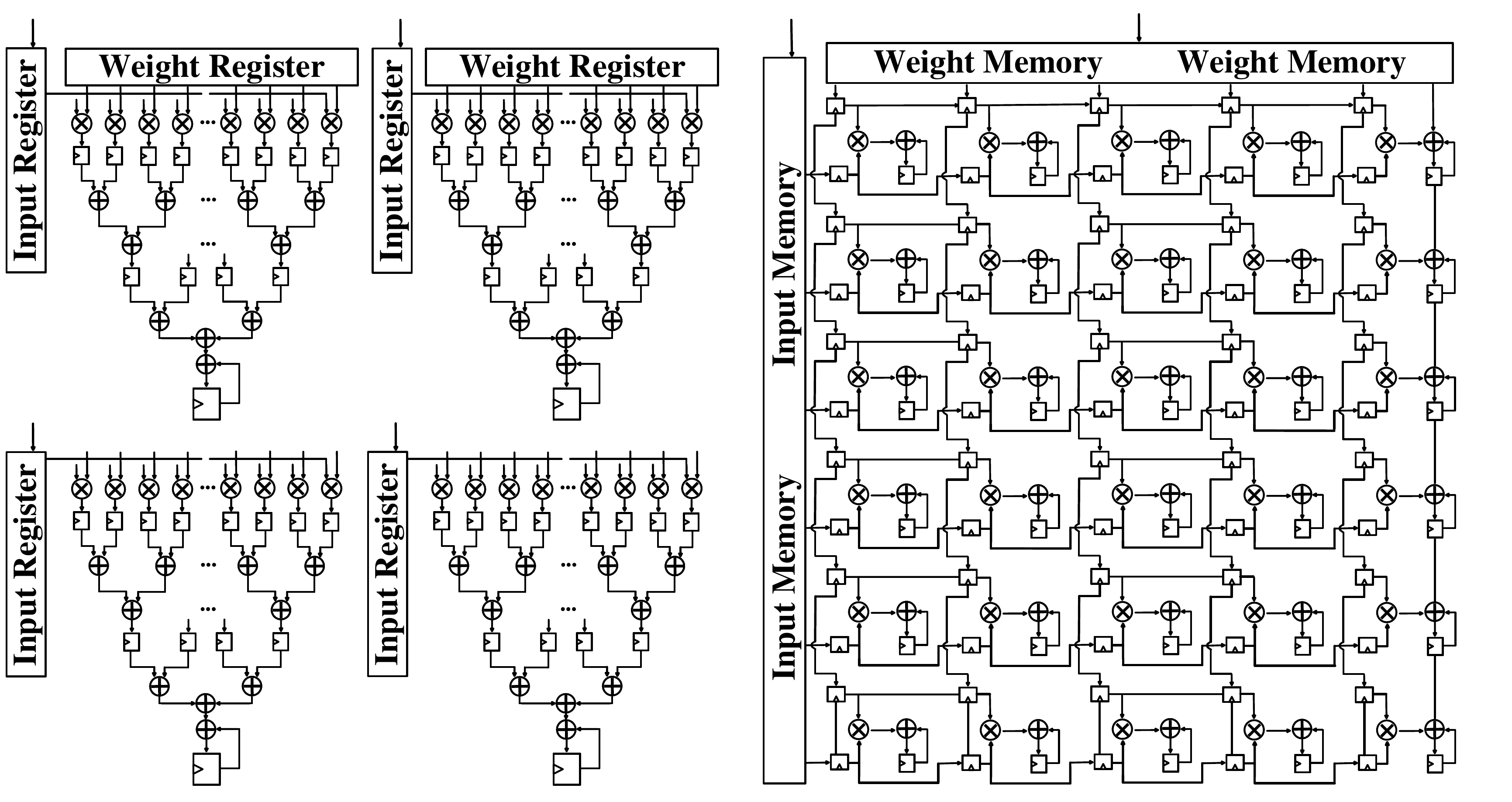}
    \caption{Microarchitectural Comparison of MAC Tree and Systolic Array}
    \Description{Limitations of SAs for Non-Linear and Reduction Operations}
    \label{fig:simd}
\end{figure}
SA and MAC-Tree architectures take different approaches to multiply-accumulate (MAC) computation. An SA is a 2D grid of simple processing elements (PEs), each performing only MACs and communicating via nearest-neighbor links. Inputs and weights are fed from array-edge buffers and propagate in a regular pattern, enabling high compute density and efficient data reuse. This simplicity makes SAs area- and energy-efficient for dense, regular matrix workloads.

By contrast, MAC Trees place greater emphasis on organizing computation around parallel multipliers and a hierarchical reduction network. As illustrated in Figure~\ref{fig:simd}, a MAC-Tree structure feeds a set of parallel multipliers from the weight registers at the top and the input registers on the left, and then progressively reduces the multiplier outputs through a multi-level adder tree to produce the final result. Unlike the regular nearest-neighbor dataflow in SA, this type of architecture incorporates both a high-fanout operand-delivery path and a hierarchical reduction path: the former distributes inputs and weights to multiple multipliers, while the latter incrementally merges multiple product terms across reduction stages. As parallelism increases, the implementation complexity of these two structures in physical layout and timing closure typically grows further, which can limit area efficiency at large scale. From an energy-efficiency perspective, MAC Trees usually require stronger vector broadcast and reduction networks, leading to higher on-chip data-movement energy. Moreover, under batched GEMM workloads, they are less able to exploit array-level data reuse, further degrading energy efficiency.

However, SA's utilization can drop substantially when the workload exhibits the following characteristic. This is a well-known classical challenge, and a large body of prior work has explored this issue~\cite{sara,ReDas,Dataflow_mirroring,ghodrati2020planaria}. Among them, prior work has also explored integrating SA with more flexible compute engines, such as CPU~\cite{ju202265nm,jeong2021rasa}, GPU~\cite{SMA}, or vector-style processing units~\cite{VSA}, to better handle irregular workloads for specific hardware scenarios. Prior flexible SA designs mainly target general shape-diverse DNN workloads, improving utilization through fine-grained reshaping and multiple dataflows at the single-array level. In contrast, SNAKE focuses on the more structured workload space of popular multi-batch LLM decode, and accordingly adopts a more targeted form of reconfigurability in array shape and dataflow. Moreover, prior work largely remains at the single-array or single-core level, whereas SNAKE further studies how such reconfigurable systolic execution should be orchestrated across multiple cores and across operators in the LLM decode setting.

We further identify 3D-stacked NMP as the natural deployment setting for reconfigurable SA for LLM decode: in lower-bandwidth platforms, decode remains memory-bound, so utilization improvements yield limited benefit, whereas the high local bandwidth of 3D-stacked NMP both makes such gains effective and enables buffer-to-compute area reallocation.

\textbf{3D-Stacked Accelerators}
To address the high bandwidth demand of LLM inference, the accelerators have evolved from bank-level in-DRAM processing designs~\cite{heo2024neupims,park2024attacc,lee20221ynm} to NMP providing higher compute capability in the logic die~\cite{yun2024duplex}. Stratum~\cite{pan2025stratum} further introduces the co-design of the 3D DRAM organization and expert activation characteristics in MoE.
Prior work mainly adopts MAC-Tree-based architectures, and the available compute capability remains mismatched with the memory bandwidth, while paying limited attention to the MNK-level shape characteristics. 

Although recent works lean toward directly integrating larger compute arrays to study thermal behavior~\cite{he2025tasa} and attention-centric LLM serving design deployment~\cite{li2026helios}, they assume a logic-die area that is difficult to reconcile with realistic HBM3-class, and even near-term HBM4-class, packages, without sufficient justification for its manufacturing and packaging feasibility. As a result, some of their conclusions do not directly apply to our setting. For example, Tasa studies thermal behavior under many-core scaling, whereas our work focuses on improving compute-area efficiency under a fixed logic-die area budget. SK Hynix~\cite{han2025near} attempts to deploy one SA under normal area budget, but they target both prefill and decode simultaneously. This work advocates a heterogeneous architecture that couples conventional XPUs with 3D-stacked NMP, where compute-intensive prefill is handled by resource-rich XPUs, while memory-dominated decode is offloaded to area-constrained NMP logic dies to better exploit their high local bandwidth. 

In contrast to prior work, which primarily optimizes 3D-stacked inference from the perspectives of DRAM organization~\cite{pan2025stratum,yun2024duplex} and operator mapping~\cite{li2025h2,li2026helios,huang2025hd}, this work rethinks the compute microarchitecture under the stringent area constraints of the logic die. As such, it is largely orthogonal to existing 3D-stacked memory optimizations and can be combined with them.

\section{KEY OBSERVATIONS}

\subsection{Motivation: The Need for Reconfigurability} 
\label{SA BASIC}

\begin{figure}[t]
    \centering
\includegraphics[width=\linewidth,trim=0.7cm 2.2cm 0.3cm 0.3cm,clip]{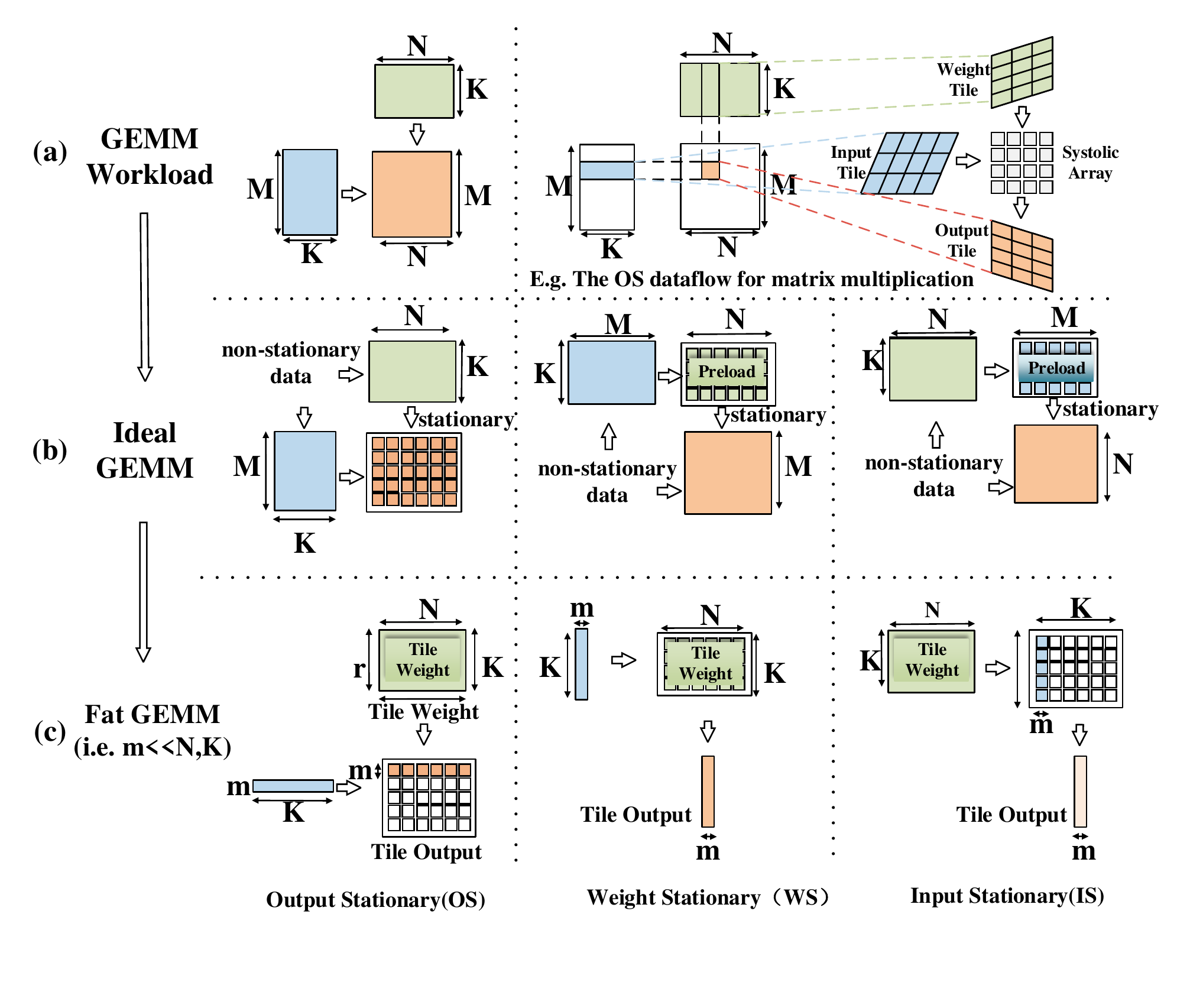}
    \caption{GEMM Abstraction and Systolic Dataflows}
    \Description{GEMM Abstraction and Systolic Dataflows}
    \label{fig:gemm workload}
\end{figure}

To unify LLM linear operators, we abstract them as GEMM $A \times B = C$, where $A \in \mathbb{R}^{M \times K}$, $B \in \mathbb{R}^{K \times N}$, and $C \in \mathbb{R}^{M \times N}$. In a 2D SA, two of $\{M,N,K\}$ are mapped to the array’s two spatial dimensions, while the remaining one is unfolded temporally. As illustrated in Figure~\ref{fig:gemm workload}(b), output-stationary (OS) maps $M$ and $N$ spatially and unfolds $K$ over time, whereas IS maps $M$ and $K$ spatially and unfolds $N$.

Since practical GEMMs usually exceed a single array’s capacity, they must be tiled and executed across multiple rounds. As illustrated in Figure~\ref{fig:gemm workload}(a), dimensions mapped to the array rows or columns require spatial tiling when they exceed array capacity, while the temporal dimension may also be segmented into multiple phases when it is too long for continuous buffered execution. For example, under OS, oversized $M$ or $N$ leads to spatial tiling, whereas oversized $K$ is split temporally. These spatial and temporal tiles, together with the fat-GEMM pattern in Figure~\ref{fig:gemm workload}(c), form the basic units for later parallel scheduling across multiple cores.

The LLM decode workloads in Figure~\ref{fig:gemm workload}(c) impose two requirements on SA reconfigurability of shape and dataflow:

First, this workload characteristic motivates array-shape reconfigurability. Decode operators are generally characterized by $M \ll N, K$, and further partitioning the already small $M$ dimension across cores is undesirable because it would incur large-weight replication with significant communication overhead~\cite{yun2024duplex,pan2025stratum}. As a result, the per-core workload usually preserves the original $M$ dimension, which itself remains dynamic across batch sizes and attention configurations, e.g., GQA or MLA. The resulting single-core decode tiles still typically satisfy $N, K \gg M$.

Conventional SA usually adopt a near-square shape for generality, since such organizations can better accommodate diverse GEMM shapes. Conventional SA also tend to employ relatively large arrays, since larger arrays usually offer higher data reuse and better area efficiency than smaller ones. Therefore, in decode the mapped $M$ dimension is often smaller than the array dimension assigned to it, leaving many PEs idle across operators and requests. Under 3D-stacked NMP, where abundant local bandwidth often shifts decode into the compute-bound regime, this utilization loss can directly translate into higher latency, while also weakening data reuse and energy efficiency.

Second, this workload characteristic also motivates dataflow reconfigurability. Since single-core decode tiles also typically satisfy $N, K \gg M$, it is sufficient to consider only OS and input-stationary (IS) while excluding weight-stationary (WS). As shown in Figure~\ref{fig:gemm workload}(c), a suitable dataflow should place one of the two large dimensions, $N$ or $K$, on the temporal dimension, so that each tile sustains longer execution, better amortizes data-loading and startup overheads, and reduce tile switching. By contrast, WS relies more on the much smaller $M$ dimension and is therefore less suitable for decode. 

Accordingly, IS is generally preferable when $N > K$, because making $N$ temporal reduces repeated re-tiling and rereading along $N$; conversely, when $K \geq N$, OS is more favorable because it unfolds $K$ temporally. To validate this first-order trend, we profile all operators of OPT~66B decode under the decode configuration detailed in Table~\ref{tab:model_configs}, at batch size 8, and separately extract the single-core tiled workloads with $N > K$ and with $N \leq K$. As shown in Figure~\ref{Impact-fig}(b), the two groups exhibit different preferred dataflows on average, while the final best choice can still vary with runtime conditions because dataflow choice affects both array cycles and memory-side stall cycles. Here, stall cycles arise when double-buffered tile refills cannot keep pace with array consumption due to insufficient effective refill bandwidth.

\subsection{Opportunity 1: High Local Bandwidth Enables Buffer-to-Compute Reallocation.}

In conventional SA constrained by off-chip bandwidth, on-chip SRAM buffers are typically large for two reasons. First, they support double buffering, so that one region supplies the current tile while another prefetches the next tile to hide memory latency. Second, they keep inputs, weights, and partial sums on chip for reuse. As a result, buffering occupies a large fraction of area. By contrast, each PE implements only simple MAC logic and is much smaller than large SRAM arrays.
\begin{figure}[t]
    \centering
\includegraphics[width=\linewidth,trim=0 0cm 0 0cm,clip]{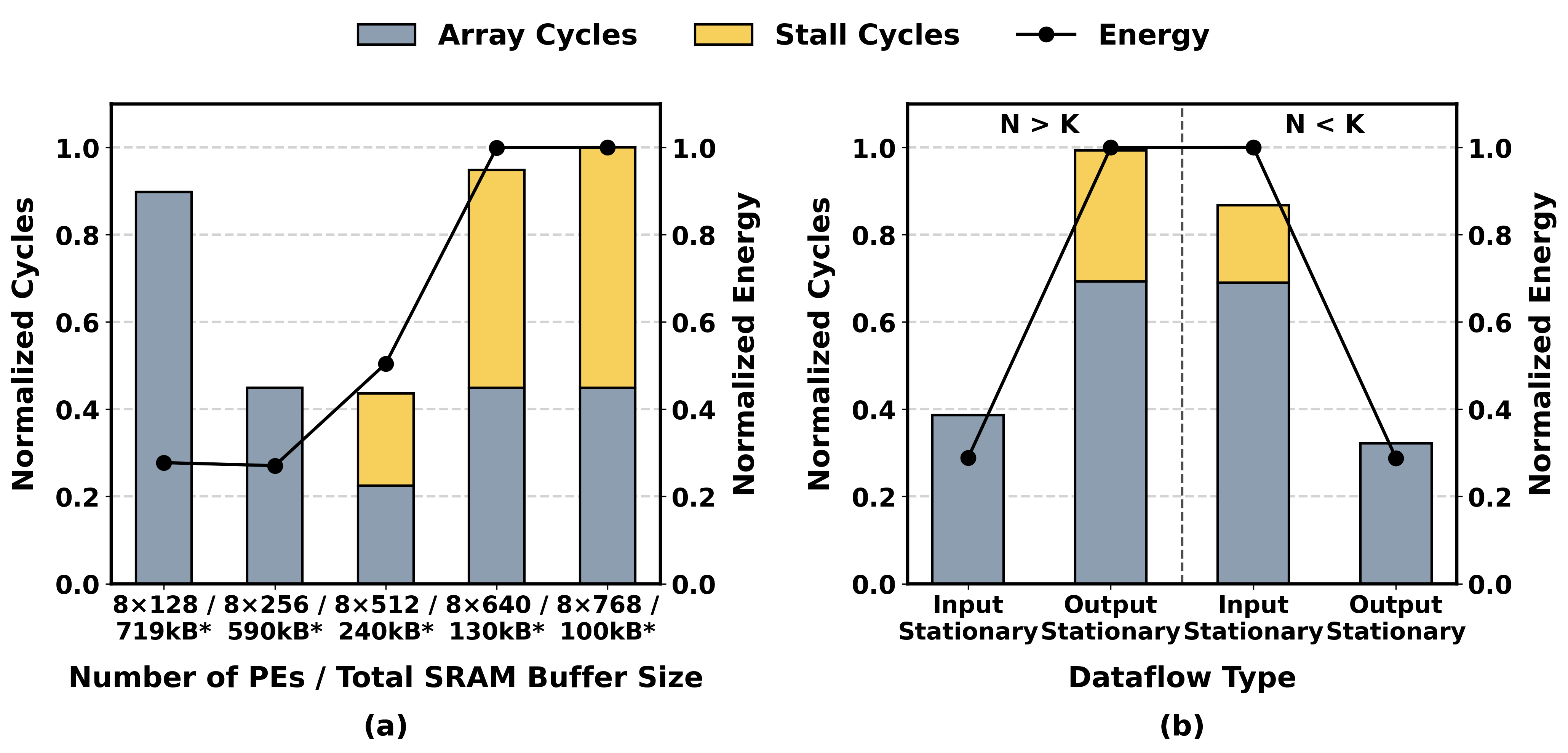}
    \caption{Impact of Buffer Allocation and Dataflow Choice on Decode Execution}
    \Description{Impact of Buffer Allocation and Dataflow Choice on Decode Execution}
    \label{Impact-fig}
\end{figure}
Under NMP, the high local bandwidth of 3D-stacked memory and the lower cost of data movement weaken these conventional reasons for large buffers. Meanwhile, as discussed earlier, decode becomes more likely to be limited by effective compute supply. To examine this opportunity, we also use OPT 66B (batch = 8), as a representative decode workload. We allocate most of the SRAM budget to the weight buffer and keep the input and output buffers small. We then keep the total area budget fixed, gradually reduce SRAM capacity, and use the reclaimed area to increase the number of PEs, thereby exploring a buffer--compute reallocation space.

Figure~\ref{Impact-fig}(a) shows the trade-off. As the PE count increases from 8 $\times$ 128 to 8 $\times$ 512, array cycles drop significantly, showing that the benefit of extra compute outweighs the loss of buffer capacity. However, when the configuration is further scaled to 8 $\times$ 640 and 8 $\times$ 768, stall cycles and energy rise sharply, indicating that the buffer has become too small to sustain efficient data supply, and the array dimensions also introduce unfavorable tile mismatch. We therefore choose an elongated 8 $\times$ 512 physical organization as the PE configuration. As shown later in Section~\ref{buffer}, batch size 8 corresponds to the highest weight-buffer demand and therefore represents the most conservative point in our analysis, while the benefit of provisioning more compute becomes more pronounced at higher batch sizes.

This result highlights two effects. First, for compute-bound decode operators, more compute units relieve the bottleneck more effectively, and the larger array can also reduce tile folds, lowering both compute and memory latency. Second, under high-bandwidth NMP, data staging and prefetching complete much faster, so large double-buffered SRAMs are less necessary. Even when a smaller buffer occasionally causes extra DRAM accesses, the penalty is much lower than in conventional off-chip-bandwidth-limited platforms. Prior work reported a similar trend. TETRIS~\cite{gao2017tetris} targets CNN/FC inference and uses an Eyeriss-based~\cite{chen2016eyeriss} accelerator in each 3D-memory vault to shift area from the global buffer to more PEs. Our work instead targets shape-diverse LLM decode and uses the reclaimed area not just for more fixed compute, but for a reconfigurable systolic substrate that adapts both array shape and dataflow to the workload.

\subsection{Opportunity 2: Unifying Vector-Style Flexibility with Systolic Efficiency.}
\label{Unifying}

\begin{figure}[t]
    \centering
\includegraphics[width=\linewidth,trim=0 0cm 0 0cm,clip]{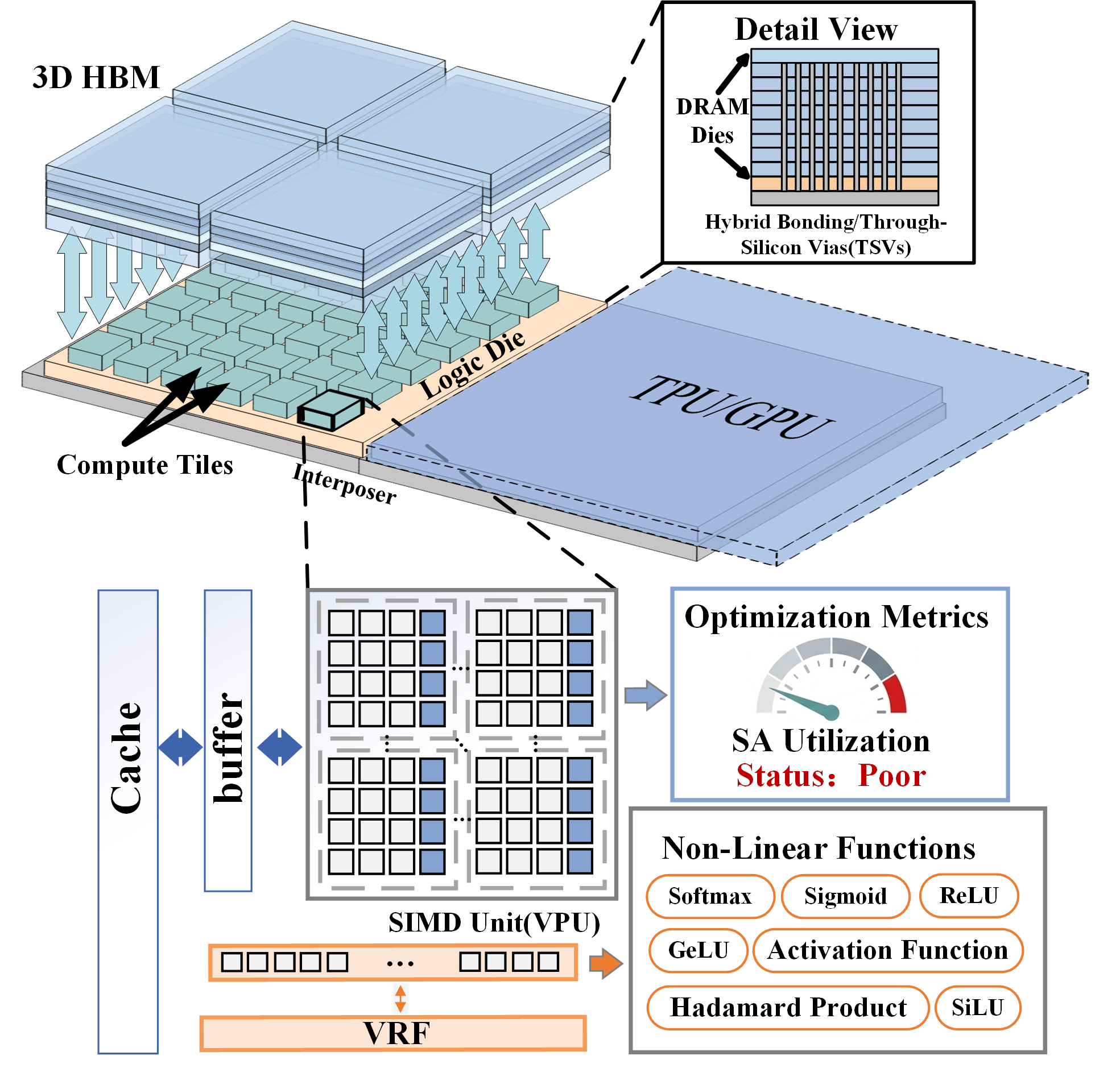}
    \caption{Vector and Systolic Core Compute Substrate for 3D-Stacked NMP.}
    \Description{Overview of a 3D-stacked NMP system architecture.}
    \label{fig:vectore-systolic}
\end{figure}

Existing 3D NMP designs follow different architectural paths~\cite{pan2025stratum,li2026helios,he2025tasa}, but they ultimately require two classes of capability at the same time. As shown in Figure~\ref{fig:vectore-systolic}, the first is high-density computation for tensor MAC operations. The second is flexible execution support for nonlinear operators, reductions, element-wise processing, and fine-grained control. The former aligns naturally with systolic execution, whereas the latter is more closely associated with vector-style execution.

In 3D-stacked NMP, the area budget on the logic die is extremely constrained. Our earlier analysis shows that decode workloads require not only high-density MAC capability, but also flexibility to accommodate dynamic variation in array shape and dataflow. This suggests that a more promising direction is not to mechanically place a vector core beside a SA, but rather to build a unified compute substrate that reuses vector-side control, buffering, and reduction support to provide vector-style flexibility together with systolic-style efficiency.

\section{ARCHITECTURE}

\begin{figure*}[t]
    \centering
    \includegraphics[width=\textwidth]{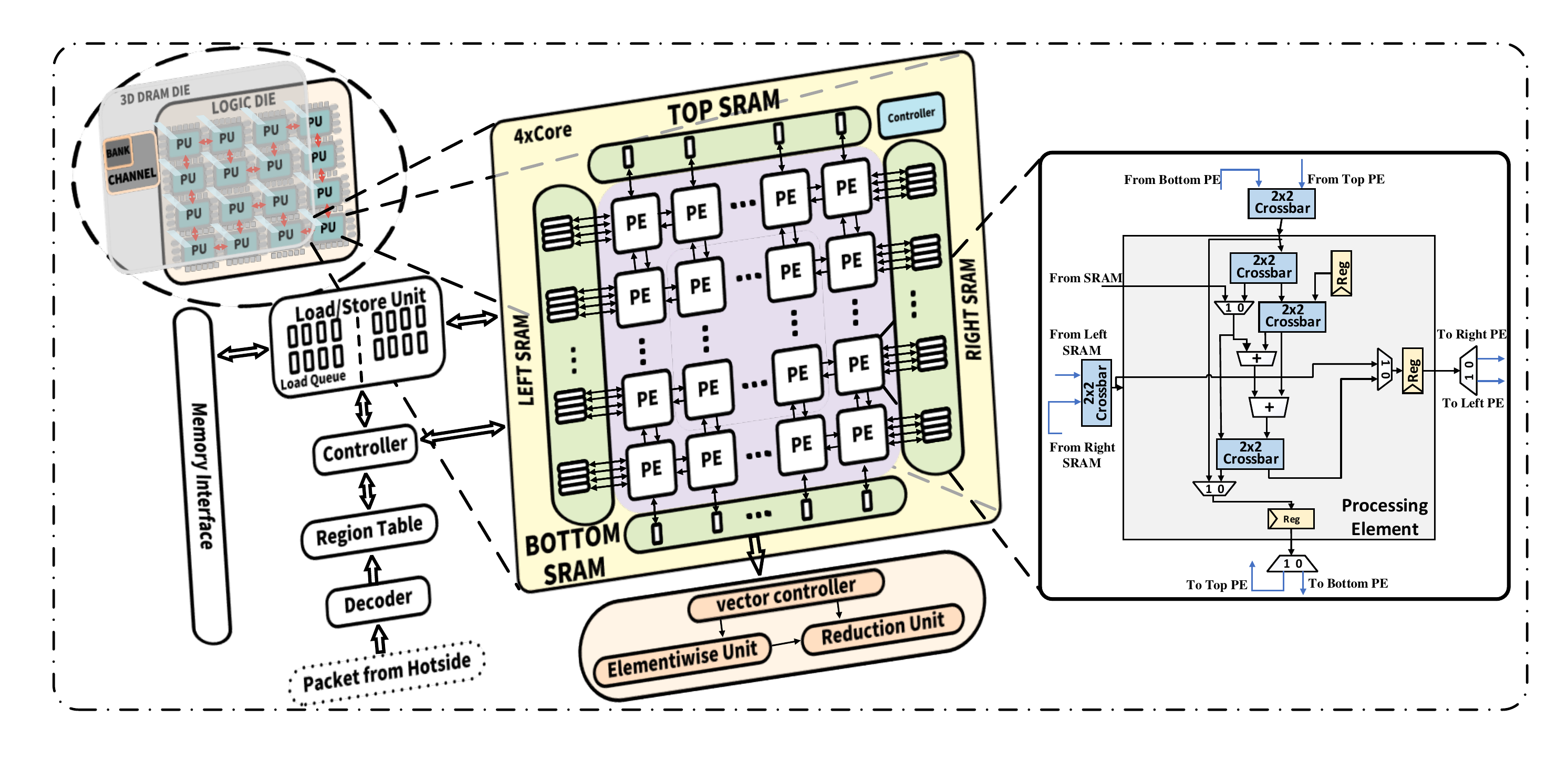}
    \caption{Architecture Overview.}
    \Description{Architecture Overview.}
    \label{fig:VISTA architecture overview}
\end{figure*}

\subsection{Overview}
Prior 3D NMP systems largely share a common stack-level template: logic-die-side compute engines are tightly coupled with stacked high-bandwidth memory. This work targets a different level of that design stack. Rather than redesigning the memory stack itself, we focus on the compute substrate on the logic die and develop an architecture that can be instantiated under similar stack-level assumptions, and the detailed reference setting is described in Section~\ref{baseline}. The external xPU remains responsible for host-side coordination and complementary execution of operators from LLM prefill. In evaluation, we instantiate this organization using an HBM3 configuration consistent with prior 3D NMP studies.

At the logic-die level, we organize the stack into 16 processing units (PUs) connected by a lightweight NoC similar to Stratum~\cite{pan2025stratum}, which is used only for coarse-grained communication such as all-reduce and all-gather. Each PU is bound to one memory channel and forms a locality-preserving compute-memory tile. Inside each PU, four compute cores are integrated. Each core is built around a 64×64 PE-based compute fabric together with local systolic buffering, while the detailed systolic-vector integration is described in Section~\ref{s-v}. 
Like prior work~\cite{yun2024duplex,pan2025stratum}, local channel access is preferred and fine-grained remote accesses are minimized. Within one PU, the four compute cores cooperatively execute the assigned local workload. To efficiently feed these cores, banks in the local channel are grouped into bank bundles for parallel tile refill. Data fetched from these bank bundles are first staged in the private systolic buffers, which provide lightweight layout reorganization and double-buffered tile supply before delivering tile-ready operand streams into the local SA.

\subsection{Core Microarchitecture}
\label{s-v}
Following the analysis in Section~\ref{Unifying}, we tightly integrate the vector core with the SA, so that the fine-grained execution substrate of the former can be reused to support fine-grained computation and reconfiguration in the latter.

\subsubsection{Array Design}

As discussed above, our design already exploits the high local memory bandwidth of 3D-stacked NMP to reduce on-chip buffering and reallocate more area to PE arrays, thereby increasing compute density under the tight logic-die area budget. However, even after reducing the overall buffer footprint, the cost of reconfigurability itself remains a major concern. In a fine-grained reconfigurable SA, multi-ported buffers are still required to support sub-array reshaping and dataflow switching under different logical mappings, and prior work has shown that they are among the dominant area costs of such designs~\cite{ReDas}. Therefore, the key objective of our array design is not only to shrink total buffer capacity, but also to further minimize the multi-port overhead needed for reconfiguration.

To make room for this capability, we reclaim part of the vector-side multi-ported buffering budget. In conventional \textit{Systolic Core + Vector Core} LLM accelerators~\cite{wu2025combating,pan2025stratum,he2025tasa}, the vector core typically relies on a heavily provisioned multi-ported local buffer to sustain reductions such as sum/max and element-wise post-processing at high standalone throughput. In our design, we do not preserve such a private buffer. Instead, we redirect part of that area to the SA side, where multi-port support more directly improves the utilization of small-$M$ decode operators mapped onto a large physical array. This trade-off is reasonable because vector-side nonlinear and reduction operators in LLM inference are typically small in scale and highly pipeline-friendly, so their latency can often be overlapped with preceding or subsequent GEMM execution~\cite{pan2025stratum}. As described later, we further tighten this integration by allowing the SA and the vector core to share a common 2-read/2-write output buffer, similar to Gemmini~\cite{genc2021gemmini}, instead of maintaining separate multi-ported storage structures as in prior 3D-stacked NMP designs~\cite{yun2024duplex,pan2025stratum}.

\begin{figure}[t]
    \centering
    \includegraphics[width=\linewidth,trim=0 0.6cm 0 0.5cm,clip]{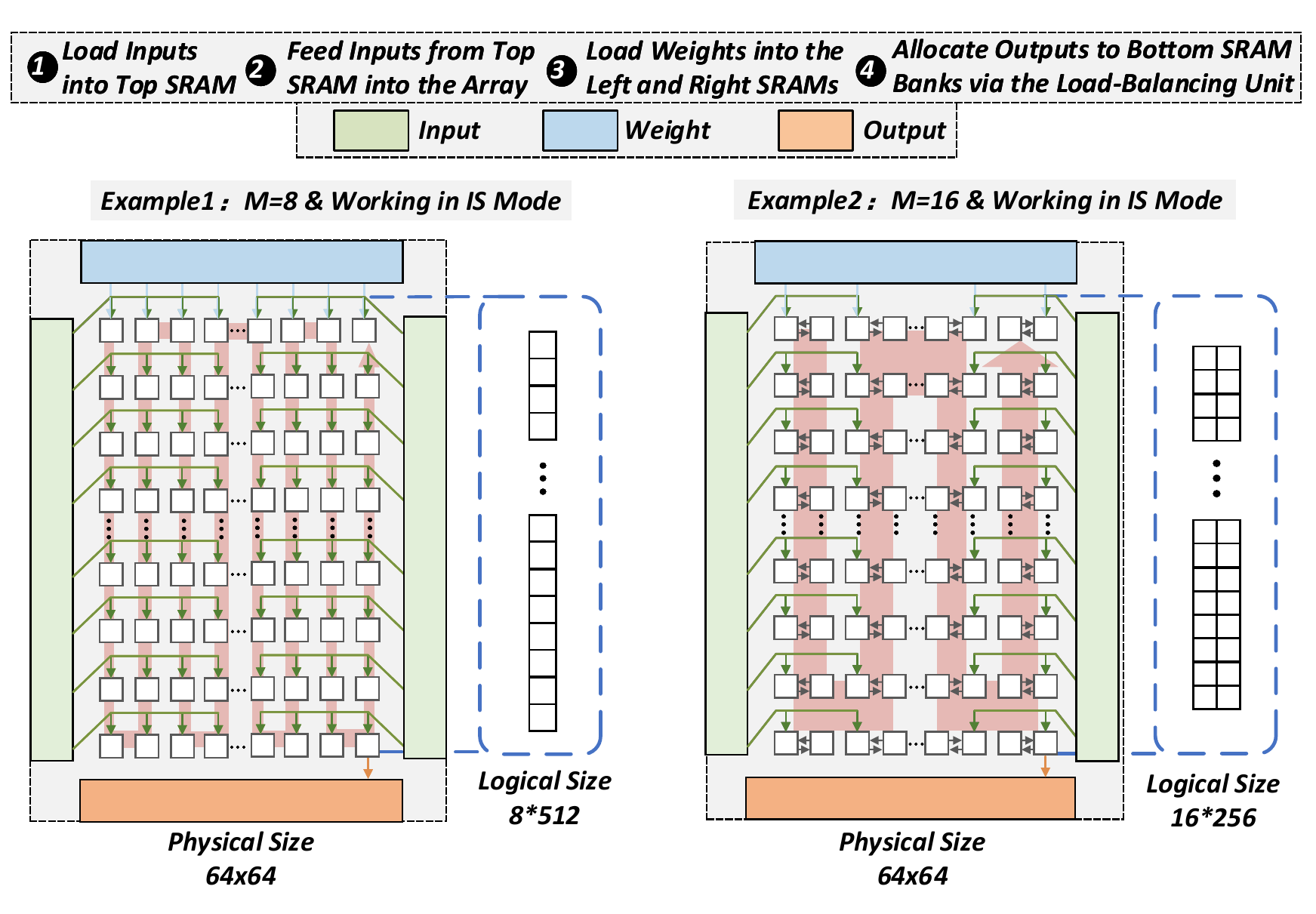}
    \caption{Serpentine Logical Array Remapping under IS Dataflow.}
    \Description{Serpentine Logical Array Remapping under IS Dataflow.}
    \label{fig:work_region}
\end{figure}

Unlike conventional SA, our design surrounds the PE fabric with four boundary buffers on the top, bottom, left, and right sides. This organization provides the structural basis for flexible remapping under different logical array shapes and dataflows. However, not all four sides require the same degree of flexibility: since fine-grained remapping is only needed along the small-$M$ dimension, multi-port support is required only for weight injection. Specifically, if the array is partitioned into $g$ logical slices, then $g$ weight-injection ports are needed in total. As shown in Figure~\ref{fig:VISTA architecture overview}, the central yellow region in the figure corresponds to a single PE array at the core level. We adopt a symmetric left-right boundary organization for the weight side, so that these ports can be distributed across the two boundaries rather than concentrated on one side. For example, when the remapping granularity is 8, only four ports need to be activated on the left boundary and four on the right boundary. This significantly reduces the per-side multi-port requirement of the weight buffer and boundary interconnect under the tight logic-die area budget of 3D-stacked NMP. Accordingly, the left and right boundary buffers can be provisioned with larger capacity, since they consistently act as the weight-side storage and injection buffers under both OS and IS. 

Inside each PE, configurable crossbars and register paths support directional data propagation, enabling the logical communication pattern to be reorganized under different sub-array shapes and dataflows. In addition, as shown in Figure~\ref{fig:VISTA architecture overview}, the rightmost zoom-in shows the microarchitecture of an individual PE. The bottom-boundary PEs introduce extra interconnect paths to support different output-propagation patterns. The reconfiguration overhead is negligible, as switching the PE mode takes only one cycle.


\subsubsection{SNAKE-Like Mapping}
On top of this low-cost multi-port organization, we propose a mapping method, termed SNAKE-like mapping, in which the dataflow propagates through the physical PE fabric along a serpentine path, much like the Snake game gradually sweeping across and filling the entire array.  Because OS and IS share the same nearest-neighbor systolic fabric and differ primarily in boundary-level operand injection and output propagation, the same physical array organization can support both dataflows. Figure~\ref{fig:work_region} illustrates this remapping under the IS dataflow. A physical $64\times64$ array serves as the fixed PE fabric. In this example, inputs are loaded into the top buffer, weights stream into the array from the left and right buffer, and the output space is distributed across different bottom SRAM banks by the SRAM load-balancing unit. Under this organization, the same physical $64\times64$ array can be remapped into different elongated logical sub-array shapes to better match operators with different small-$M$ dimensions. When $M=8$, the entire physical array is reorganized into a logical $8\times512$ sub-array. In other words, the original two-dimensional set of 4096 PEs is traversed in a SNAKE-like manner and concatenated into a longer logical computation path, allowing each logical output stream to propagate through the full PE fabric and accumulate continuously. Likewise, when $M=16$, the same physical array is reorganized into a logical $16\times256$ structure, enabling more parallel logical output streams while preserving a long effective computation path. Although our minimum reconfiguration granularity is 8 rather than 1, this limitation mainly affects a small subset of GEMV-like decode operators with extremely small $M$ (e.g., $M=1$), which cannot be perfectly matched by our logical sub-array shapes. This design choice is acceptable because such cases tend to become memory-bound, where execution is dominated by data-supply stalls rather than peak compute occupancy. In such cases, the residual utilization loss has limited impact on end-to-end performance.
\subsubsection{Vector Core Design}
The vector core is organized around the output buffer of SA, rather than using a separate private SRAM. In this way, the same storage structure serves both as the writeback space for SA results and as the input/intermediate buffer for subsequent vector operations such as softmax, normalization, and other element-wise processing. To support this interaction, output buffer is implemented as a banked 2-read/2-write structure. The SA can write results back to the buffer, while the vector core reads them out for post-processing. A lightweight arbitration logic coordinates array writeback, array readback, and vector-side accesses according to access type and bank conflicts. This organization preserves limited overlap between tensor and vector execution when accesses do not conflict, while keeping the vector-side storage overhead small under the tight logic-die area budget.

\subsubsection{Control Logic Design}

The vector core originally employs fine-grained control, and we extend this control logic to also manage the systolic mode. As shown in the middle of Figure~\ref{fig:VISTA architecture overview}, after the Decoder parses the operator descriptor, the controller centrally dispatches the corresponding commands to the LSU and to the RTAB. 

\textbf{Decoder.}
The Decoder receives task packets from the host side and parses them into internal execution descriptors for both the systolic core and the vector core. For systolic execution, each matrix-multiply instruction is further split into pipelined sub-stages---Weight Load, Feed First/Second, Drain---to overlap execution and mask PE idle cycles. Control signals flow with operands, minimizing both hardware overhead and reconfiguration latency, similar in spirit to RASA's pipelining optimizations~\cite{jeong2021rasa}.

\textbf{Load/Store Unit (LSU).} 
The LSU manages on-chip SRAM at bank granularity, handling mode selection, address allocation, while hiding storage latency. Under different dataflows, it assigns matrix tiles to different boundary SRAMs and schedules their load/feed order accordingly, so that the same physical core can realize different logical operand movements without changing the PE fabric. For vector execution, it also supports the movement of intermediate tiles between the output buffer and vector-side operations.

\textbf{Region Table (RTAB).}
The RTAB records the spatial boundaries of each working region together with its corresponding SRAM allocation results under different dataflow. During execution, it continuously tracks the readiness, running, and completion states of all regions, enabling dynamic management of multiple logical sub-arrays.

\section{MULTI-PU SCHEDULING VIA SPATIAL AND SPATIO-TEMPORAL PARTITIONING}

\textbf{a) Intra-Operator Multi-PU Scheduling.}

In this part, we perform multi-core scheduling at the PU granularity. Each PU contains four local reconfigurable SA cores that cooperatively execute the assigned linear operators under a unified systolic dataflow for each operator. As discussed in Section~\ref{SA BASIC}, the preferred single-core systolic dataflow depends on the relative size of $N$ and $K$. We therefore turn to the view of MNK-level partitioning.


First, we do not partition the $M$ dimension across PU as described above, and keep the $M$ dimension inside each PU to determine logical array shape and only design multi-PU scheduling over the two dominant dimensions, $N$ and $K$. Under IS and OS, these two dimensions naturally serve as the spatial and temporal dimensions in different ways. As a result, the multi-PU design space reduces to four partitioning modes.

\textbf{1) IS-S: IS with pure spatial partitioning.}
In this mode, IS is used as the inter-PU dataflow, and only spatial partitioning is applied across PU. Different PU mainly split the $K$ dimension in space, while each PU locally processes its tiles along the $N$ time dimension.

\textbf{2) IS-ST: IS with spatio-temporal partitioning.}
This mode extends IS-S by further partitioning the time dimension. Besides splitting the $K$ dimension in space, it also divides the time dimension $N$ into multiple stages or time blocks.

\textbf{3) OS-S and 4) OS-ST.}
Similarly, OS-S partitions the $N$ dimension in space and advances along the $K$ dimension in time. OS-ST further partitions the $K$ dimension into multiple time blocks based on OS-S.

\begin{figure}[t]
    \centering
\includegraphics[width=\linewidth,trim=0 0cm 0 0cm,clip]{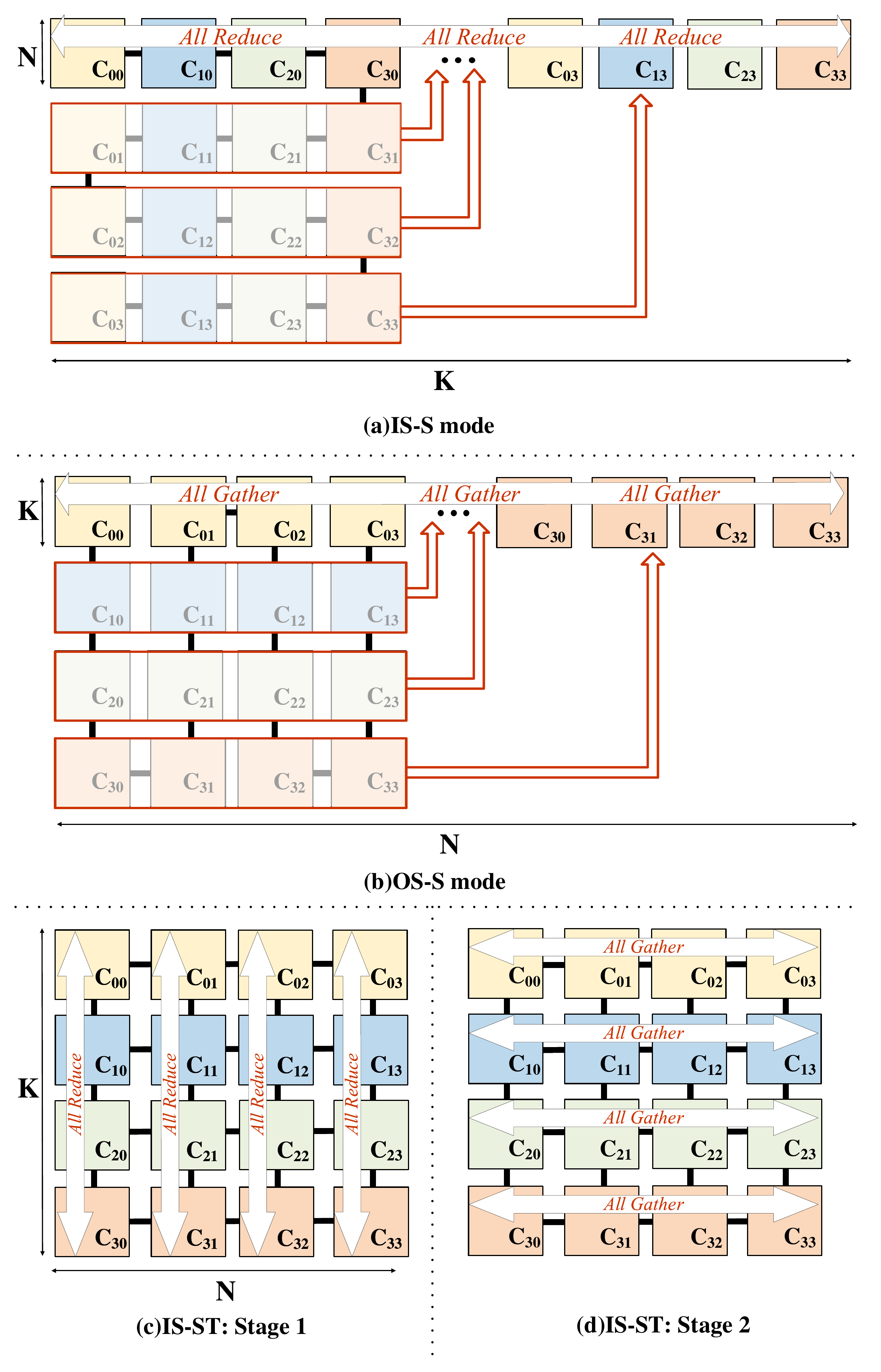}
    \caption{Logical Inter-PU Organizations for Representative Partitioning Modes}
    \Description{Logical Inter-PU Organizations for the Four Partitioning Modes}
    \label{fig:isos}
\end{figure}

This lightweight interconnect design follows from the regular \(N\)- and \(K\)-dimension partitioning, which limits inter-PU communication to a few collectives and keeps the structure simple and area-efficient. As shown in Figure~\ref{fig:isos}, the same 16 PU can be organized in two simple logical ways using interconnect: a SNAKE-like $1\times16$ chain, similar to the traversal described above, or a regular $4\times4$ mesh, similar to a conventional array interconnect. The $1\times16$ view is used when one dimension is divided by 16 across the 16 cores, corresponding to \textit{IS-S} and \textit{OS-S}. The $4\times4$ view is used when both dimensions are divided by 4 across PU, corresponding to \textit{IS-ST} and \textit{OS-ST}.

\textbf{b) Operator-Specific Scheduling Analysis.}
These four modes define the MNK-level scheduling space for the major linear kernels in decode. Among these scheduling candidates, the final mode is selected based on the overall operator-specific considerations described below. We first decide which operators are scheduled within this space and which are better handled separately. We also analyze tile-level overlap between these linear operators and the following nonlinear stages to reduce end-to-end latency.

For the $QK$ and $AV$ head-level operators in attention, the computation of one head is usually small, and thus cannot fully hide the memory access latency from 3D-stacked memory. Therefore, we follow a strategy similar to Stratum~\cite{pan2025stratum}: we mainly use head-level parallelism (partition the $M$ dimension), map different heads to different PU, and improve utilization by interleaving the linear stage and the Softmax/reduction stage of different heads within the same group.

For the remaining decode operators that account for most of the execution time, such as projection operators and expert FFNs in MoE layers, we treat them as independent GEMM/GEMV kernels and analyze their dataflow choices under the above multi-PU spatial/spatio-temporal framework.

\begin{figure}[t]
    \centering
\includegraphics[width=\linewidth]{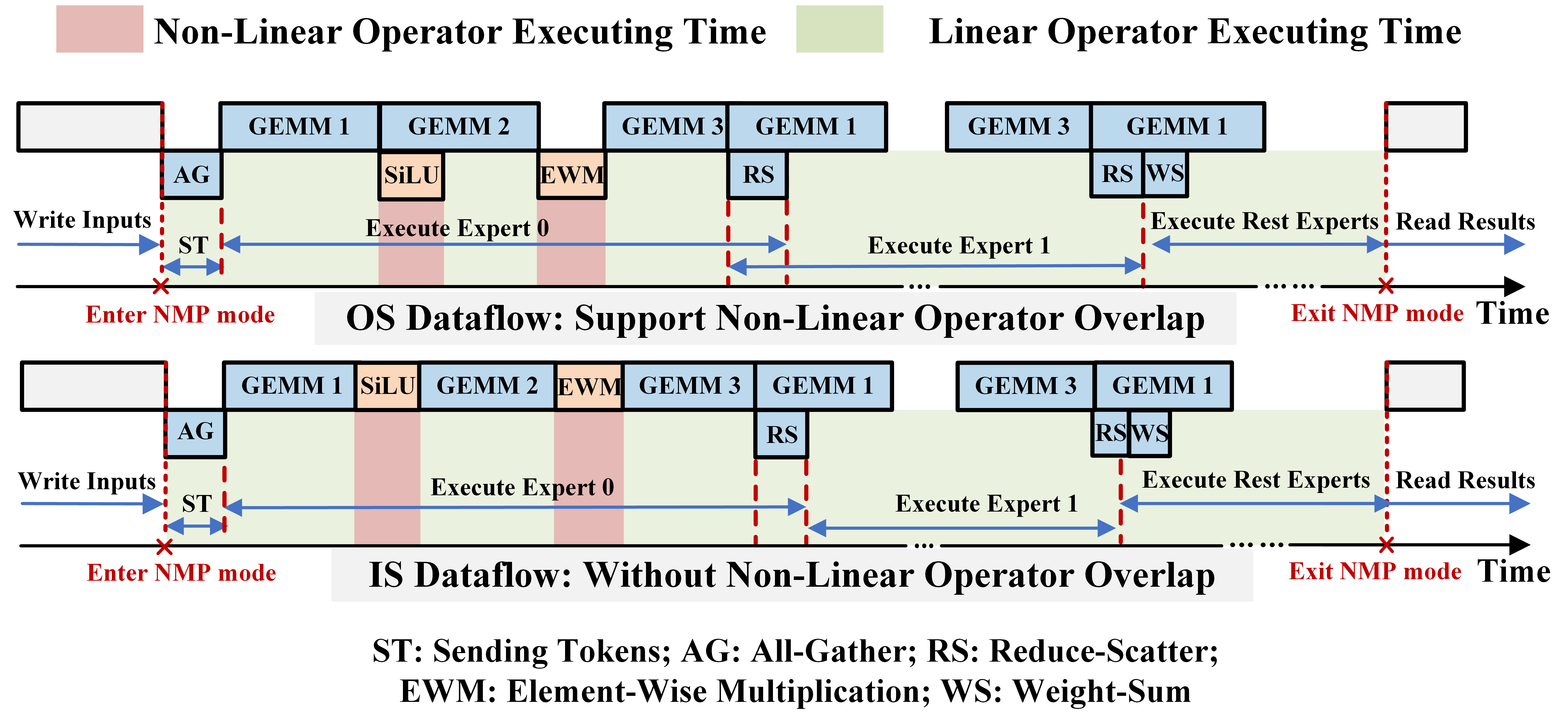}
    \caption{A Representative MoE-Expert Example of Linear--Nonlinear Overlap under OS and IS Dataflows.}
    \Description{A diagram showing Overlap.}
    \label{fig:Overlap}
\end{figure}

In general, OS dataflow is more favorable for direct tile-level overlap between linear and nonlinear stages, because an output tile can be consumed by the following nonlinear operator as soon as its in-array reduction is completed. By contrast, under IS dataflow, such direct overlap is usually weaker, since an output tile often becomes available only after accumulation along the temporal dimension is finished. Below we instantiate this framework with a representative decode scheduling policy in Fig.~\ref{fig:Overlap} for MoE layer in DeepSeek 236B (batch = 8). Still, IS can remain favorable when it shortens the critical GEMM latency, and it may also exploit overlap across independent operator branches. In practice, the realizable overlap further depends on whether the following stage is tile-foldable, as well as on the communication pattern (e.g., all-gather vs.\ all-reduce) and buffer-capacity constraints. 

Overall, this yields a compact and deployment-friendly scheduling space, where each operator only needs to evaluate four candidate strategies. For a given LLM, our simulator performs this lightweight search and selects the best partitioning strategy for each operator and assembles the corresponding best scheduling combination for the full network.

\section{EVALUATION}
\subsection{Experimental Setup}

\subsubsection{Benchmark}
We follow Helios~\cite{li2026helios} in benchmark selection and use OPT~\cite{zhang2022opt}, LLaMA3 ~\cite{grattafiori2024llama3}, Mixtral ~\cite{jiang2024mixtral}, Qwen3~\cite{qwen2025qwen3}, and DeepSeek~\cite{deepseek2024v3}. These models cover both dense and MoE LLMs. We use model-specific input/output length settings for decode and serving evaluation. Detailed configurations are summarized in Table~\ref{tab:model_configs}. We follow prior work~\cite{yun2024duplex} and model MoE expert activation using a uniform routing distribution during evaluation.

\begin{figure*}[t]
    \centering
    \includegraphics[width=\textwidth, trim={0 0.cm 0 0}, clip]{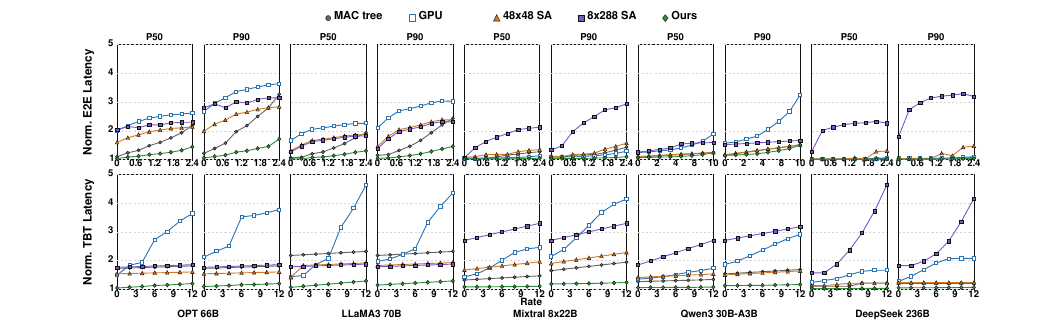}
    \caption{Normalized Serving Latency Under Different Request Rates}
    \Description{servinglatency}
    \label{fig:serving_perf}
\end{figure*}

\subsubsection{Baseline}
\label{baseline}

\textbf{System template.} Since we focus on compute-substrate comparison rather than full-system redesign, we adopt Stratum~\cite{pan2025stratum} as a common HBM3-class 3D-stacked NMP template. It represents one of the most aggressive and well-documented prior designs in terms of internal DRAM bandwidth and active logic-die compute budget. Accordingly, we reuse its memory-system assumptions, including DRAM organization, timing, and energy parameters, and fix the effective DRAM bandwidth at 24~TB/s, the midpoint of Stratum's reported range.
    
\textbf{Baseline compute substrates.} Under this common system template, we instantiate a MAC-Tree design as the representative baseline compute substrate. To further isolate the benefit of reconfigurability, we also implement two fixed-shape SA baselines: a square $48\times48$ array and a long $8\times288$ array. Under the same 3D-NMP logic-die constraints, these designs represent practical fixed-shape alternatives without the microarchitectural flexibility and area optimization enabled by our design. We also use GPU as the baseline.
    
\textbf{Area-normalized comparison.} Since our architecture adopts a compute hierarchy structurally similar to Stratum, we use the same number of PUs under the same logic-die area budget for all designs, enabling a direct area-normalized comparison. All other system-level assumptions are kept unchanged.
    
\textbf{Frequency assumption.} The fixed-shape baselines are assumed to operate at up to 1~GHz, whereas our reconfigurable design targets 800~MHz due to the additional routing constraints introduced by reconfigurability. Therefore, our comparison is area-normalized and implementation-aware, rather than iso-frequency.

\subsubsection{Modeling}
Using SystemVerilog, we implement our architecture by modifying PLENA~\cite{wu2025combating}, a representative LLM SA design, and adapt its original flattened systolic substrate into a reconfigurable SA for the 3D-stacked near-memory setting configured in IEEE 754 FP16 under the 7 nm ASAP7 predictive PDK~\cite{clark2016asap7}. The local buffer on the logic die is implemented as SRAM macros, modeled using FinCACTI~\cite{shafaei2014fincacti} and calibrated with publicly available SRAM specifications~\cite{chang201712}. Dynamic energy is estimated from the post-synthesis gate-level netlist with switching activity annotated from simulation.

For kernel-level performance modeling, we build on Scale-Sim v3~\cite{raj2025scalesimv3}, which integrates the Ramulator~\cite{kim2015ramulator} memory model, and extend it to capture SA execution in a 3D-NMP setting, including multi-PU interconnection and vector core. 
\begin{figure}[t]
    \centering
    \includegraphics[width=\linewidth,trim=0 0.3cm 0 0.8cm,clip]{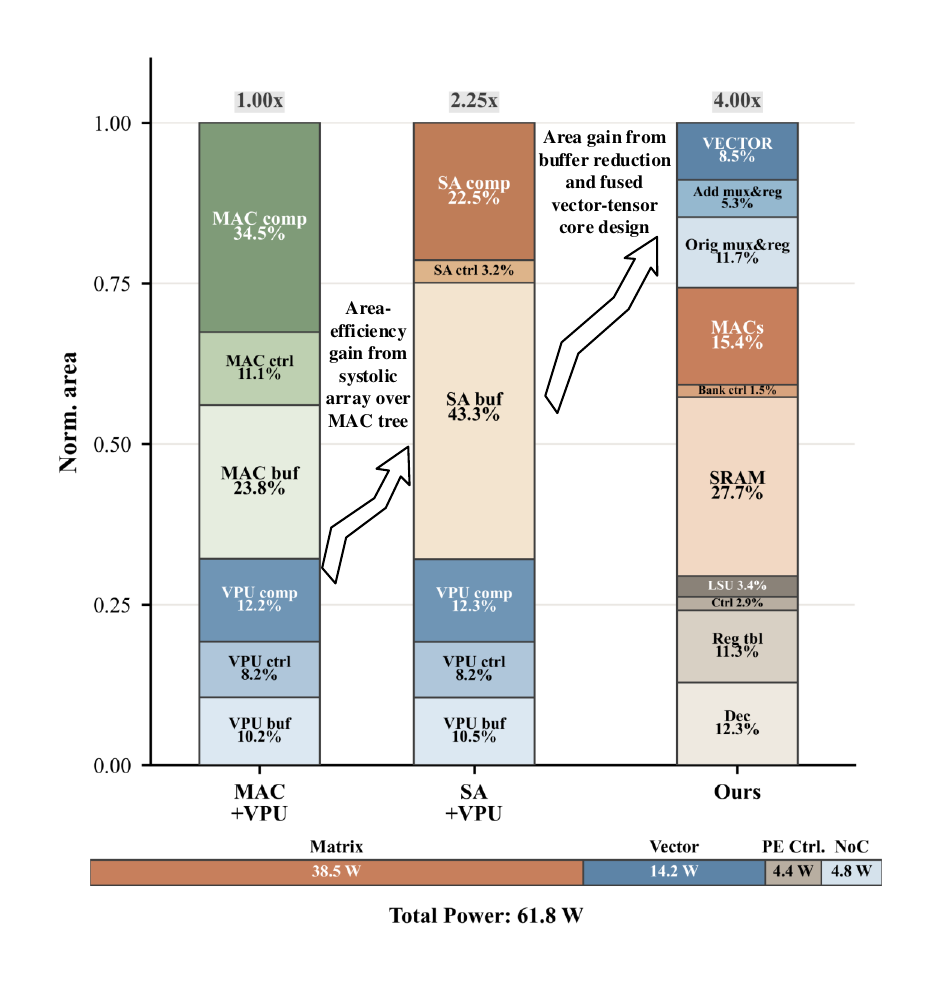}
    \caption{Normalized PU-level Area and Power Breakdown}
    \Description{Normalized area breakdown and compute-area efficiency}
    \label{fig:Area-Breakdown}
\end{figure}
For end-to-end serving evaluation, as well as GPU and MAC-Tree baseline characterization, we build on Duplex's system-level serving framework~\cite{yun2024duplex}, including Poisson-based request injection under varying arrival rates, continuous batching, and latency accounting. We use NVIDIA H100 as the common prefill engine across all compared systems, and also use H100-only decoding as one baseline. Within the Duplex-based simulator, we retain its internal GPU and NVLink models and incorporate a Stratum-configured MAC-Tree backend for comparison. We evaluate all models on an 8-device system with tensor parallelism degree TP=8. Although our reconfigurable SA can potentially support more flexible scheduling for MoE expert layer in 3D NMP~\cite{huang2025hd}, we intentionally retain TP for these layers in this work as in~\cite{pan2025stratum,yun2024duplex}.

\begin{table}[t]
\centering
\caption{Architectural configurations of evaluated models.}
\label{tab:model_configs}
\small
\setlength{\tabcolsep}{3pt}
\renewcommand{\arraystretch}{1.0}
\begin{tabular}{lcccl}
\toprule
Model & L & (H, F) & (Q, KV) & Configuration \\
\midrule
OPT 66B        & 64 & (9216, 36864) & (72, 72)   & Dense, MHA \\
LLaMA3 70B     & 80 & (8192, 28672) & (64, 8)    & Dense, GQA \\
Mixtral 8$\times$22B & 56 & (6144, 16384) & (48, 8)    & MoE, E=8, top-2 \\
Qwen3 30B-A3B  & 48 & (2048, 768)   & (32, 4)    & MoE, E=128, top-8 \\
DeepSeek 236B  & 60 & (5120, 1536)  & (128, 128) & MoE, E=160, top-8, MLA \\
\bottomrule
\end{tabular}
\end{table}

\subsection{Area Breakdown and Thermal Characterization}

Figure~\ref{fig:Area-Breakdown} compares the normalized area breakdown and compute-area efficiency of three RTL-implemented designs at single-PU granularity under the same PU-level area budget of 2.35~mm$^2$. For fairness, each design uses the largest feasible compute-unit configuration under this budget: the MAC-Tree baseline adopts a $16\times16\times16$ organization, the conventional SA baselines use $4\times48\times48$, and our design uses $4\times64\times64$. 

Replacing the MAC-Tree engine with SA already brings a clear improvement: SA + Vector Core reaches 2.25$\times$ higher compute-area efficiency. This gain mainly comes from the structure of the SA with higher compute density.
Our design further improves area efficiency in two ways mentioned before, and achieves 4.00$\times$ compute-area efficiency over the MAC Tree. 
\begin{figure*}[t]
    \centering
    \includegraphics[width=\textwidth]{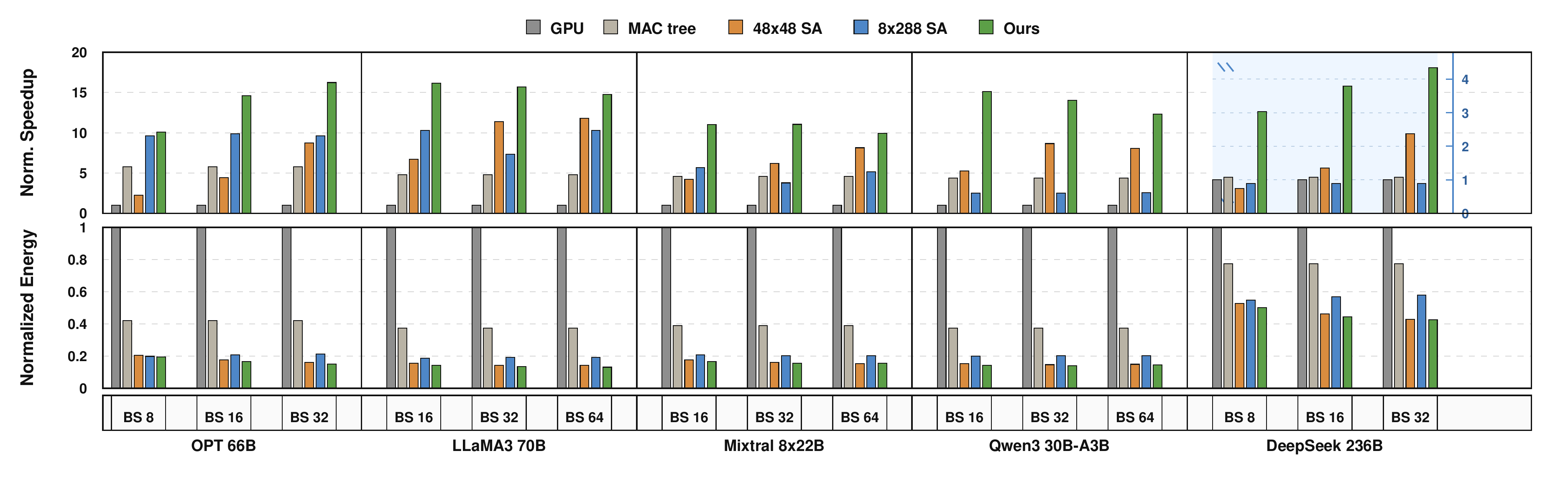}
    \caption{Normalized Decode Performance and Energy Efficiency}
    \Description{decodingnormmetrics}
    \label{decoding_norm_metrics}
\end{figure*}
From the area breakdown, additional Muxes and Regs at the PE level for SA reconfigurability account for 6.0\% of the total area, which is offset by the area saved. Notably, although our buffers now include multi-port SRAMs to support fine-grained reconfigurability, the total buffering-related area still decreases from 53.6\% in \textsc{SA + Vector Core} to 28.1\% in \textsc{Our Work}, and this released area is primarily reallocated to compute units. The vector-core share also decreases to 8.8\%, with part of it further reallocated to fine-grained control logic and compute units.

Following Stratum, we use HotSpot~\cite{zhang2015hotspot} to evaluate the 3D thermal behavior under the same package and cooling assumptions, and define the logic-die power budget as the maximum logic power that keeps the peak temperature below 85°C~\cite{han2021power}. For practicality, we reduce the effective DRAM bandwidth from 30.34 TB/s to 24 TB/s, which lowers DRAM power and leaves more thermal budget for the logic die. We also use a lower logic-die frequency, 800 MHz, which further reduces dynamic power. As a result, the allowable logic-die power budget is about 62 W. Figure~\ref{fig:Area-Breakdown} shows the peak-performance power breakdown of the logic die under the thermal operating point described above. The total logic-die power is 61.8 W, including 38.5 W for Matrix unit, 14.2 W for Vector unit, 4.4 W for PE control, and 4.8 W for NoC.

\subsection{Decode Performance Evaluation}
\label{decode}

Prior work~\cite{he2025tasa} has established that thermal dissipation at the logic die is the primary bottleneck in 3D-stacked systems. Therefore, this work focuses specifically on comparing the energy consumption of the logic die. As shown in Fig.~\ref{decoding_norm_metrics}, compared with GPU, our design achieves an average of $11.47\times$ speedup and $5.74\times$ higher energy efficiency. Compared with the MAC-Tree baseline, our design achieves an average of $2.90\times$ speedup and $2.40\times$ higher energy efficiency. These gains come from both a denser compute organization under the tight logic-die area budget and better workload matching through reconfigurable dataflow and Multi-PU scheduling. Meanwhile, in addition to lower clock frequency and DRAM bandwidth, by reducing unnecessary on-chip SRAM accesses and shortening execution time, our design also improves energy efficiency.

Further comparison with the two fixed-shape SA baselines shows that reconfigurability itself is a key source of the gain. Compared with the $48\times48$ SA baseline, our design achieves an average of $2.33\times$ speedup and $1.05\times$ higher energy efficiency; compared with the $8\times288$ SA baseline, it achieves $3.00\times$ speedup and $1.31\times$ higher energy efficiency on average. Although a fixed-shape SA may achieve better local reuse for some operators, the reconfigurable SA can select a more suitable array shape for each operator, thereby improving both throughput and memory-access energy efficiency.

\subsection{Serving Performance Evaluation}

Fig.~\ref{fig:serving_perf} presents the latency results of five models under two serving scenarios: the top row reports end-to-end (E2E) latency under full serving, while the bottom row reports time-between-token (TBT) latency under decoding stress tests.

Fig.~\ref{fig:serving_perf} shows normalized serving latency under different request rates for 8K-input and 1K-output requests. The x-axis is the normalized request rate, and the y-axis reports normalized latency, including both end-to-end (E2E) latency and time-between-tokens (TBT) latency, with our design as the baseline.

Specifically, the GPU baseline exhibits substantially higher latency. As shown in the figure, GPU E2E latency is typically around $1.5\times$--$3.0\times$ that of our design, while its TBT latency usually reaches about $1.5\times$--$4.0\times$. MAC-Tree is generally the closest baseline to our design, with E2E latency mostly around $1.1\times$--$2.3\times$ that of our design and TBT latency mostly around $1.3\times$--$2.2\times$. As the request rate increases and the effective decode batch becomes larger, the area-efficiency advantage of SA becomes more consistent. Since our design can further adapt both array shape and mapping strategy to the workload, it gradually widens the gap over MAC-Tree.

For fixed-shape SA, the $48\times48$ SA consistently performs better than the $8\times288$ SA, but both remain inferior to our design. In particular, the $48\times48$ SA still shows about $1.1\times$--$2.4\times$ E2E latency and $1.1\times$--$2.2\times$ TBT latency of our design, while the $8\times288$ SA is often worse, with TBT latency reaching about $1.5\times$--$4.5\times$. This suggests that a near-square array is more robust than a highly elongated one under serving workloads. Still, both fixed-shape designs suffer from workload-shape sensitivity, whereas our reconfigurable array can sustain higher utilization across more diverse decode workloads.

\subsection{Multi-PU Scheduling Analysis}

\begin{figure}[t]
    \centering
\includegraphics[width=\linewidth,trim=0cm 0cm 0cm 0cm,clip]{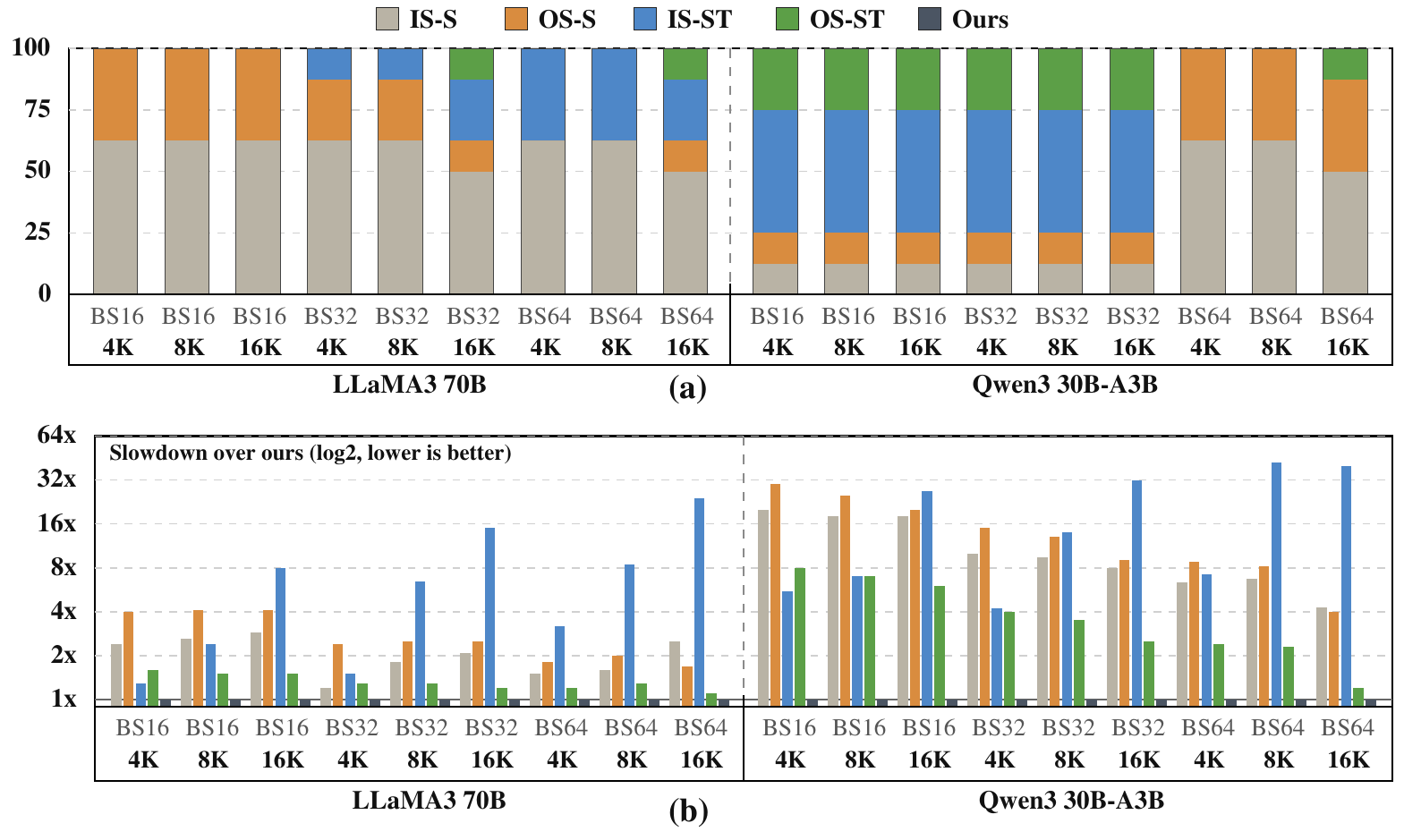}
    \caption{Scheduling Mode Selection and Fixed-Mode Slowdown}
    \Description{(a) Per-operator Partitioning Mode Distribution (b) Performance Comparison: Fixed Scheduling Mode vs. Per-Layer Flexible Scheduling}
    \label{strategy}
\end{figure}

With logical array shape determined, Figure~\ref{strategy} (a) shows the distribution of per-operator selected multi-PU scheduling modes across all operators under different batch sizes and input sequence lengths. We include one dense model (LLaMA3~70B) and one MoE model (Qwen3~30B-A3B) for comparison. For the dense model, strategy selection is highly concentrated, with IS-S dominating (59.7\%), followed by OS-S (25.0\%), IS-ST (12.5\%), and OS-ST (2.8\%). In contrast, the MoE model exhibits a more balanced distribution across IS-ST (33.3\%), IS-S (27.8\%), OS-S (20.8\%), and OS-ST (18.1\%). This illustrates that optimal partitioning strategies vary significantly across operators, making it challenging for any single fixed strategy to achieve consistently high utilization.  

Figure~\ref{strategy} (b) compares the slowdown of using one fixed scheduling mode for all operators against our per-operator flexible scheduler. For LLaMA3~70B, the best fixed strategy (OS-ST) still incurs a slowdown of 1.04$\times$–1.56$\times$ relative to the per-operator scheduler. The impact is more pronounced for Qwen3~30B-A3B, where the best fixed policy experiences a slowdown of 1.18$\times$–6.43$\times$. These results highlight that a per-operator, workload-adaptive strategy is essential to approach optimal decode performance across diverse operators and sequence lengths.

\subsection{Array-Shape and Buffer Trade-offs}
\label{buffer}
\begin{figure}[t]
    \centering
\includegraphics[width=\linewidth,trim=0cm 0cm 0cm 0cm,clip]{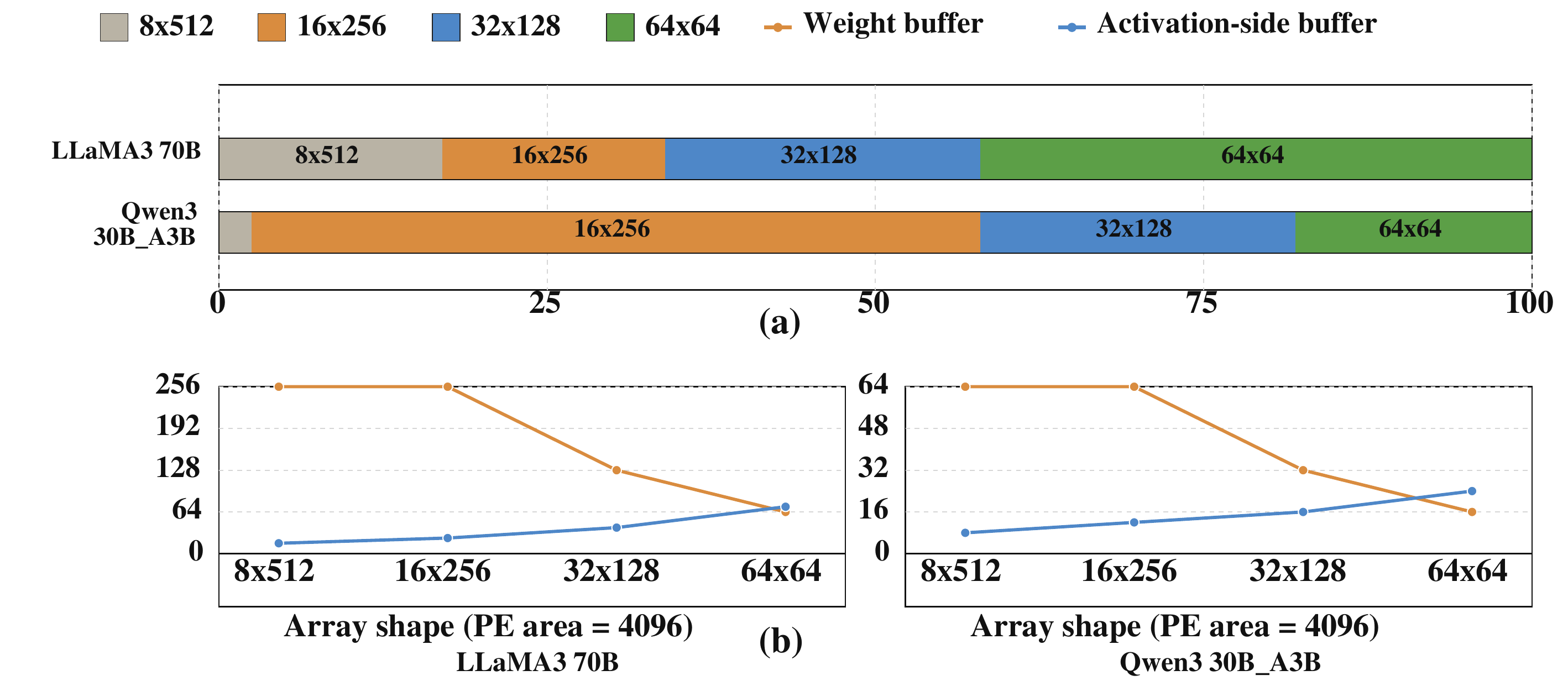}
    \caption{(a) Array-Shape Demand Across Models (b) Buffer Requirements Across Array Shapes}
    \Description{Array-Shape Distribution and Minimum Buffer Requirementsn}
    \label{Array-Shape}
\end{figure}

Figure~\ref{Array-Shape} (a) shows the distribution of selected array shapes across batch sizes from 8 to 64 for two representative models under a fixed PE budget. Figure~\ref{Array-Shape} (b) reports, for each array shape, the minimum weight buffer and activation-side buffer capacities required to sustain stall-free tiled execution. activation-side buffer is output under IS, while it is the input buffers under OS. We focus on these two type of buffers because they dominate the buffering demand, while the remaining operand stays stationary in the array and needs little buffer space. First, since the $M$ dimension of decode operators is largely determined by batch size, the preferred logical array shape also shifts accordingly, although not in a strictly one-to-one manner across all operators. Second, as the array becomes narrower and less elongated, the demand on the filter buffer decreases, while the demand on the activation-side buffer increases, revealing a clear trade-off between the two. 

\section{CONCLUSION}
Overall, this work shows that co-designing a reconfigurable systolic array and workload-aware multi-core scheduling effectively converts the high local bandwidth of 3D-stacked NMP into better LLM decode performance and energy efficiency.


\bibliographystyle{ACM-Reference-Format}
\bibliography{sample-base}

\end{document}